\documentstyle[12pt,epsfig]{article}
 1
\def\as{\alpha_{\rm S}}

\def\citenum#1{{\def\@cite##1##2{##1}\cite{#1}}}
\def\citea#1{\@cite{#1}{}}

\def\as{\alpha_{\rm S}}

\def\g{\gamma}

\def\s{\sigma}

\def\({\left(}
\def\){\right)}

\def\citenum#1{{\def\@cite##1##2{##1}\cite{#1}}}
\def\citea#1{\@cite{#1}{}}

\def\l1vt{\vec{l_{1\perp}}}

\def\rt{r_{\perp}}
\def\bt{b_{\perp}}
\def\rt2{$r^2_{\perp}$}
\def\bt2{$b^2_t$}

\def\jol1{$J_0(\,l_{1\perp}\,r_{\perp}\,)$}

\def\citea#1{\@cite{#1}{}}

\def\beq{\begin{equation}}
\def\eeq{\end{equation}}
\def\bea{\begin{eqnarray}}
\def\eea{\end{eqnarray}}

\def\eq#1{{eq.~(\ref{#1})}}

\def\bbbz{{\mathchoice {\hbox{$\sf\textstyle Z\kern-0.4em Z$}}
{\hbox{$\sf\textstyle Z\kern-0.4em Z$}}
{\hbox{$\sf\scriptstyle Z\kern-0.3em Z$}}
{\hbox{$\sf\scriptscriptstyle Z\kern-0.2em Z$}}}}
\def\l{\lambda}
\relax
\begin{document}
\begin{titlepage}
\noindent
\begin{flushright}
 August  1998\\
DESY 98 - 118\\
TAUP 2521/98\\
\end{flushright}
\begin{center}

{\Large\bf{25 YEARS WITH THE POMERON}}\\[9ex]

{ \Large \bf{ Eugene  Levin ${}^{\dagger}$
\footnotetext{$^{\dagger}$ E-mail: leving@post.tau.ac.il} 
}} \\[1.5ex]
{\it School of Physics and Astronomy}\\
{\it Raymond and Beverly Sackler Faculty of Exact Science}\\
{\it Tel Aviv University, Tel Aviv, 69978, ISRAEL}\\
{\it and}\\
{\it Theory Department, Petersburg Nuclear Physics Institute}\\
{\it 188350, Gatchina, St. Petersburg, RUSSIA}\\
{\it and}\\
{\it DESY Theory Group}\\
{\it 22603, Hamburg, GERMANY}\\
[3.5ex]
\end{center}
\vspace{1cm}
\centerline{ \em Paper for ``Gribov Memorial Volume" , Heavy Ion
Physics,Acta Physica Hungarica}
\vspace{1cm}
{\large \bf Abstract:}
 This is a report on  my  25 years activity in understanding of the 
Pomeron structure. Since I was involved in the Pomeron business
moreless from the beginning, I hope, that this report shows the 
development of the main ideas in their historical perspective from the 
first enthusiastic attempts to find a simple solution to understanding 
of
the complexity and difficulty of the problem.

 In other words, this is a 
story about  a young guy who wanted to understand everything in high 
energy interaction, who did his best but who is still in the beginning,
 but who has not lost his temper and considers the Pomeron structure as 
the beautiful and difficult problem, which deserves his time and  
efforts to be solved.  
\end{titlepage}
\section{Instead of introduction.}
I started to be involved in the Pomeron problem in early 70's 
not because I felt that 
this was an interesting problem for me but rather because everybody 
around worked with this problem.

 It was a heroic time in our department ( theory department in St.
Petersburg (Leningrad) Nuclear Physics Institute ) 
when
we worked as one team under the leadership of Prof. Gribov. He was
 for us, young guys,  not 
only a dominant leader but a respectable teacher. I took seriously his 
words: `` Genya, it seems to me that you are a smart guy.  I am sure, 
you will be able to do something more reasonable than this quark stuff."
So I decided to try and, frankly speaking, behind my decision  there was 
another reason. I felt that I could not develop my calculation skill, 
doing the additive quark model. However, I must admit,  very soon I 
got a deep interest in the Pomeron problem, so deep  that I decide to
present here all my ups and downs in the attempts to attack this 
problem.
\section{Reggeon Calculus.}
My first remark is that in 70's to find the high energy asymptotic was
a highly priority job. During the last five years I have traveled a lot
around the globe and I have found that it is very difficult to explain 
for young physicist why it was so. However, 25 years ago the common 
believe was that the analytical property of the scattering amplitude 
together with its asymptotic will give us the complete and 
selfconsistent theory for the strong interaction. The formula of our 
hope was very simple:
\beq \label{1}
 ``Analyticity,\, Unitarity\,\,  and\,\, Crossing\,\,+\,\,Asymptotic\,\,
=
\,\,Theory \,\,of\, 
\,
Everything"
\eeq
It is clear that to find the second part of the above formula was a 
great challenge for theorists. The main ingredients of our attempt to 
approach the asymptotic behaviour of the scattering amplitude were and 
are the Reggeons which came up as the solution of the first puzzle: the 
contradiction between the experimental observation 
 of the hadrons ( resonances ) with spin ($j$) higher than $1$
 and the steep energy behaviour of the scattering amplitude ( $A$) due to 
exchange such resonances.

 Indeed, the exchange of a hadron with spin 
$j$ leads to $ A\,\propto\,s^j$, where $s$ is the invariant energy of 
the reaction. For $j\,>\,1$  $A$ exceeds the Froissart boundary 
\cite{FRST} which has been proven from unitarity, analyticity and 
crossing symmetry \cite{MAR}. The solution to this puzzle was to assume 
that all resonance with certain quantum numbers could be described by one
function of the resonance mass ($t$) ( trajectory $\alpha(t)$ ). At $t = 
M^2_R$, where $M_R$ is the  mass of resonance with spin $j$,
  $\alpha(t= M^2_R) = j $. In the scattering kinematic region $ t\,<\,0$
 and $\alpha(t)$ should be smaller than 1. In this case $s^{\alpha(t)}$ 
does not exceed the Froissart boundary . The concept of Reggeons is still 
the only solution to the first puzzle that have been found during  25
years of the  development of out theoretical approach.

However, it turns out that experimental trajectories give the value of 
$\alpha(t=0)$ which is well below 1.  It means that, if the asymptotic is
 defined by only Reggeon contribution, the total cross section shall fall 
down as a power of energy. Therefore, we got the second puzzle: in the 
Reggeon approach the total cross section should decrease as a function 
of energy while experimentally it is constant at high energy if not 
rising. The only way out of this puzzle was to assume that there is a 
new Reggeon ( Pomeron)  with the trajectory $\alpha_P (t = 0 ) \,=\,1$. 

It turns out that even this strong assumption cannot give us a simple 
solution for the asymptotic \cite{GRIB1}. The basic idea was that we 
can build the effective theory for high energy interaction assuming
 the Pomeron and taking into account the interaction of the Pomerons with
colliding particle and between them. V.N. Gribov found the Lagrangian
for such effective theory \cite{GRIB1} and started the approach which 
was called Gribov Reggeon Technique or Gribov Reggeon Calculus.
 A bit later
  A.A.Migdal and V.N. Gribov
made the first attempt to solve this theory  \cite{MG}.

I entered to the game just at this stage and Gribov, Migdal and me in 
Refs.\cite{GML1} tried to find the answer to the 
following question:``What could be a feedback to the high energy behaviour 
of the 
exchange of normal Reggeons with $\alpha(0) \,<\,1$ ( so called 
secondary trajectories or Reggeons ) from their interactions with the 
Pomeron?".  It was nice time to remember because we understood a lot
and found a beautiful form of the Lagrangian for the interaction of the
secondary Reggeons with the Pomeron, especially for so called  fermion 
Reggeons. In one particular case, when the secondary trajectory does not 
depend on $t$, we gave the exact solution to the problem \cite{GML1} 
which has been used in the solid state physics for more reliable 
physical applications. However, the physical result of this study for me 
was rather destructive, because we found that even sufficiently weak 
interaction between the  Pomerons led to strong effect for the secondary 
Reggeons changing crucially their power behaviour at high energy. On the
other hand, experimentally, in all reactions where we have only Reggeon 
exchange the cross section has a beautiful power behaviour.  As far as I 
know it is still an open problem and more microscopic approach such as 
QCD have not led to a solution of this puzzle. I consider this problem 
as a great challenge for QCD and for all theorists who got be involved 
in this high energy business.

The Reggeon Calculus is an effective theory for high energy interaction 
in which the transversal and longitudinal degrees of freedom are treated 
in a different way. By now, we have understood that this a general 
property for any theory at high energy.  V.A. Kudryavtzev, A.A. 
Schipakin and me in Refs.\cite{LKS}  gave the first 
example how this separation of  transversal and longitudinal degrees of
freedom simplify the treatment of the particles with high spin at high 
energy. We generalized the Reggeon Calculus for the case of particles 
with spin and found a number of predictions in Reggeon approach for
the polarization at elastic and inelastic processes. Since only the 
general properties of the Reggeon Calculus have been used, our  way to
take into account the spin of particles at high energy is still alive 
and widely used in the phenomenology of high energy interaction.

In the beginning of this section I have mentioned that the Reggeon 
Calculus was not a pure theoretical invention but rather some way of 
compromise between the general properties of our amplitude and the 
existing experimental data. We considered as a very important job to
 check our prediction with the experimental data to get a feedback to 
enrich our theoretical approach. To do this we created in 1972 KHOLERA
( KHOze,LEvin, Ryskin and Asimov ) collaboration. We had basically two 
goals: (i)  to check how well (or bad ) we could describe the current 
experimental data using everything that we had learned theoretically 
and (ii) to provide a community service since our experimentalist had
started  to work at CERN on elastic high energy scattering. 

We did our best \cite{K}
  and learned several lessons:
\begin{enumerate}
\item  We are able to model to experimental rise of the total cross
section, 
assuming that the Pomeron has still intercept $\alpha(0)\,=\,1$ if we 
include the Pomeron interactions;

\item  The diffraction dissociation process gives more restrictable 
information on the asymptotic behaviour of the theory than the total
or/and elastic cross section;

\item  We can provide the description of all processes assuming the
additive 
quark model and applying the Reggeon analysis to quark - quark 
scattering amplitude.
\end{enumerate}
We summarized our understanding in the lectures at our Winter School
\cite{KREV} which were translated to English. I have learned also a 
personal lesson which was and is my guide: the experimental data cannot 
specify a theory for us but can be used only to verify our theoretical 
picture for high energy interaction.

In recent years Gotsman, Maor and me in Refs.\cite{GLM}
 have repeated the KHOLERA approach with new 
assumption on the Pomeron structure, namely with the Pomeron intercept
$\alpha_P(0)\,=\,1 \,+\,\epsilon$ which naturally follows from the QCD 
approach \cite{LN}. We derived the same conclusions as have been mentioned
above, 
but , in some sense, with the opposite sign: the intercept of the 
Pomeron bigger than one does not mean  a violation of the Froissart 
boundary if the interaction between Pomerons has been taken into 
account.

  All such phenomenological approaches incorporate a lot of the 
model assumptions and can  be discussed with a lot of reservations, but 
they are the best illustration to the point that we have to understand 
better the Pomeron structure in theory on more microscopic level than in 
the Reggeon Calculus, to develop the selfconsistent  effective theory 
for
the high energy interaction.

\section{Multiperipheral Model.}

The first attack on the microscopic structure of the Pomeron was 
undertaken by us, M.G. Ryskin and me, in the framework of so called 
multiperipheral model.  The idea of this approach is very simple. The 
Pomeron has been introduced to describe the behaviour of the total cross 
section with energy. Let us try to study the total cross section itself 
to understand its energy behaviour using the optical theorem, namely
\beq \label{2}
\sigma_{tot}\,\,=\,\,\sum^{\infty}_{1} \,\int\,| A ( 2 \rightarrow n) 
|^2\,d \Phi_n\,\,,,
\eeq 
where $M(2 \rightarrow n )$ is the amplitude of the production of $n$ 
particles in the final state with the phase space $ d \Phi_n$.

At first sight we need to solve even a  more complicated 
problem,namely,
to find the amplitude for multiparticle production, but a hope was 
that we would need only general properties of such an amplitude to 
understand the principle feature of the energy behaviour of the total 
cross section. Indeed, even before we started to approach this problem
 Amati,Fubini and Stangellini \cite{AFS} as well as Ter-Martirosian 
\cite{TER} have found that \eq{1} leads to power energy behaviour
$ \s\,\propto\,s^{\alpha_P(0)\,-\,1}$ if we assume that all produced particles 
have average transverse momenta  which do not depend on the value of 
energy ( so called multiperipheral kinematic). However, the first 
attempt to evaluate the value of $\alpha_P \,-\,1 $ was discouraging
 and led to $\alpha_P \,-\,1  \,\approx\,- 0.7$. 

Our first question that we asked ourselves starting this project was to 
understand what more detail properties of hadron production we needed to
build the total cross section which did not depend on energy ( or 
better to say $\alpha_P \,-\,1 \,\ll\,1$ ). The second question was to 
understand  what really we were calculating: the Pomeron or the whole mess of 
the correct asymptotic behaviour which included the complicated Pomeron 
interaction. Our approach was based on two main ideas:
\begin{enumerate}
\item  The Pomeron interaction is small and can be neglected at
sufficiently 
high  energies, where $G_{3P}\,\ln s \,<\,1$ ( $G_{3P} $ is triple 
Pomeron coupling here ). The experimental data supported this assumption
since they gave $G_{3P}\,\approx \,0.1$\,\,;

\item  For the amplitude of $n$-particle production we can use the
Veneziano
model \cite{VEN}, which was ( and is ) am example of the solution to our
first equation.  In this model the hadron interaction was described
as creation and decays of the resonances in such a way that the sum over 
all resonances reproduced the correct Reggeon asymptotic with the only 
one shortcoming: there was no the Pomeron in this model. However, the 
last point it was just that we needed to avoid the double counting in 
our calculation.
\end{enumerate}
As a result of calculations we found \cite{MM1} that:

\begin{enumerate}
\item   In a natural fashion we are able to 
get the constant total cross 
section at high energy;

\item The process of the multiparticle production has two stages: (i) the 
production of resonance with small fraction of the direct produced pions
and (ii) the decay of resonances, which gives the final distribution of 
pions;

\item The typical distances in the first stage of the production process 
( resonance production ) are rather small ( typical transverse momenta 
 are large ) of the order $\frac{1}{ 0.5 - 1 \,GeV^2}$. The small observed 
transverse momentum of pion ( $p_{\pi \,t}\,\propto 
\,\frac{1}{m_{\pi}}$ ) is the result of the sequent resonance decays;

\item The cross section of hadron production with large transverse
momentum 
($p_t$) has scaling behaviour, namely,  
$\s\,\,=\,\,\frac{1}{p^8_t}\,F(\frac{2 p_t}{\sqrt{s}})$;

\item The Pomeron has broader multiplicity distribution than the Poisson 
one for the produced pions due to the resonance decays;

\item The rapidity correlation between pions with like-charge ( for 
example, for $\pi^- \pi^-$ ) has an exponential like fall down at large 
values of the difference in rapidity ($\Delta y = y_1 - y_2$)
$$R(y_1,y_2) \,=\,\frac{\frac{d \s}{\s\, d y_1 dy_2}}{ \frac{d \s}{\s\, d y_1}
\frac{d \s}{\s\, d y_2}}\,\,-\,\,1\,\,=\,\,R(0)\,e^{-\,\frac{|\Delta y 
|}{L}}$$
with the correlation length $L\,\approx\,1$.
\end{enumerate}
All these properties had not been known at that time but were confirmed 
experimentally later on.

\section{Multiparticle Production.}
The beauty of the Pomeron approach to high energy interaction was and, 
I believe, still is the possibility to describe in the unique pattern 
both  the 
elastic or/and semielastic processes and the processes of the 
multiparticle production. The theoretical basis of this description is 
the Abramovsky, Gribov and Kancheli cutting rules (AGK)\cite{AGK}
 which recover 
the interrelation  between both kind of processes if we know the Pomeron 
structure or in other words if we know for what kind of the 
multiparticle production processes is responsible the Pomeron exchange.

Having in hand a sufficiently reliable model for the Pomeron structure, 
the next natural step for us, M.G.Ryskin and me, was to check how the 
Pomeron interaction would reveal itself in the processes of 
multiparticle production. We did this in a number of papers ( see 
Refs.\cite{MM2} ) and our 
results could be summarized as follows:
\begin{enumerate}
\item The Pomeron interaction lead to long range rapidity correlation in 
the multiparticle process and the correlation function could be given in 
the form:
$$
R(y_1,y_2) \,=\,\frac{\frac{d \s}{\s\, d y_1 dy_2}}{ \frac{d \s}{\s\, d y_1}
\frac{d \s}{\s\, d y_2}}\,\,-\,\,1\,\,=\,\,R(0)\,e^{-\,\frac{|\Delta 
y |}{L}}\,\,+\,\,2\,\frac{\s_2}{\s_{tot}}\,\,,
$$
where $\s_2$ is the contribution to the total cross section the exchange 
of two Pomerons;

\item In the semi-inclusive processes where we measure the number of 
produced particle the long range part of the correlation is negligible;

\item The Pomeron interaction gives the sizable part of Bose-Einstein 
correlation which is the most dominant contribution for BE correlations 
in hadron-nucleus collisions (see Ref.\cite{CK} );

\item The Pomeron interaction leads the the KNO scaling in the
multiplicity
distributions which will be broken out only at the ultra high energies.
\end{enumerate}
Summarizing our efforts to understand the Pomeron structure in the 
framework of the multiperipheral model, I think, that we demonstrated the 
possibility of  such an approach and ability of this approach, taking into 
account the Pomeron interaction, to describe available at that time 
experimental data. We consider as the shortcoming of our approach the 
sufficient complexity of it and the lack of the theoretical basis. 
However, two results which came as outcome of our approach we took with 
us for future investigations: the two stage of the processes of 
multiparticle production and rather small distances essential for these 
processes.

\section{Partons, time - space structure of interaction and the 
death of the Pomeron approach.}
Looking around for a theory which could help us to understand the scattering 
amplitude at sufficiently small distances  we naturally found ourselves 
among supporters and activists of the parton model suggested by 
Feyman\cite{FE},Bjorken\cite{BJ} and Gribov\cite{GR}. This model 
proposed to look at a  fast hadron as a  system of noninteracting 
point-like particle-partons. These partons interact in normal way but 
with unknown ( at that time ) Lagrangian while the distribution of 
partons in a hadron has a definite value of the transverse momentum.
At that time, such an approach looked for us more general that the 
specific field theory and more suitable to match with our 
multiperipheral approach for the Pomeron.

Naturally, what we did first we translated our multiperipheral result into 
parton language and found even more argument for the hypothesis that the 
mean transverse momentum ($q_t$)  of partons large enough: $
q_t\,>\,1\,GeV$ \cite{P1}.

 The second, what we did, was to reconsider 
our multiperipheral result for the high $p_t$ hadron production in the 
parton model. We found 
\cite{P2} that in the parton model this cross section 
can be written trough the cross section of 
interaction of a parton from one hadron (say 1)  with a parton from 
another (say 2) and through the convolution of the two parton densities 
or
probabilities to find partons 1 and 2 in the hadrons 1 and 2, respectively.
The power behaviour of the cross section is determinated by the parton - 
parton cross section at angle about 90 degree. We found also that in 
the parton approach unlike the multiperipheral one each parton should 
produce the jet of hadrons with specific properties that have been 
studied by us.

The third, we found a simple interpretation of the AGK cutting rules in 
the framework of the parton model \cite{P3}, which gave us an 
understanding that the AGK cutting rules is a general property of any 
field theory.  

The parton model for us was a way to clarify the space - time picture of 
the hadron interaction at high energy \cite{P4}. Especially important 
for us was  paper \cite{P6} in which we developed the space-time picture
for hadron nucleus interaction, understood the physical meaning of the 
Glauber approach in the parton model and found the solution to the 
hadron-nucleus interaction in the parton approach. Unfortunately, this 
paper was done at the very end of the parton era and it passed unnoticed
by high energy community, but I consider this paper as the best of mine 
on the parton approach and the ideas and methods, developed in this 
paper I have used in the new attack on the Pomeron structure in QCD.

Just during this period we, M.G. Ryskin and me, realized that the 
Reggeon Calculus and the Pomeron approach is deadly sick. We did not 
publish anything on this subject, throwing in the waste-basin all our 
efforts to save this approach, but this fight was our way to QCD. 
Actually, the death of the attempt to build the effective theory 
starting from the Pomeron as the Regge pole has a definite date - 1975.
In 1975 McCoy and T.T. Wu \cite{TTWU} and Matinyan and Sedrokyan 
\cite{MS} showed that in the wide class of the field theories the 
exchange of the two Pomeron did not give the correct asymptotic at high 
energy. We realized very quickly that the new diagrams which have been 
suggested in Refs.\cite{TTWU} \cite{MS} have the correct space-time 
structure in the parton approach and therefore, the effective theory 
based on the Pomeron contribution and on the interaction between 
Pomerons has ceased to exist. We have to start from the beginning, 
looking for new ideas for the effective theory at high energy.
\section{DIS at low $\mathbf x$ \,\,( 20 years ago ).}
 Fortunately for us, L.V. Gribov, M.G. Ryskin and me,
 Prof. Gribov asked us a question which triggered 
our thinking in a right direction. The question was: ``what happened with 
high energy asymptotic in the deep inelastic scattering ( DIS ) where we know 
the evolution equation and, therefore, we have a solid theoretical 
basis." It took several years to find the answer. Let me list here what 
we found \cite{GLR1}{\cite{GLRREV1}\cite{LRREV}\cite{LLREV}:

{\bf 1.} The DGLAP evolution equations \cite{GLAP} give correct
asymptotic
for 
the deep inelastic structure function ( $F_2(x,Q^2)$, where $Q^2$ is the 
virtuality of photon and $x$ is Bjorken variable, the energy of 
collision is equal $s\,=\,\frac{Q^2}{x}$ ) for all values of $x$ such 
as $\ln \frac{1}{x}\,\ll\,\ln^2 Q^2$. This asymptotic leads to increase 
of $F_2$ mainly due to the growth of the gluon density ($x G(x,Q^2)$)
 in a target
\beq \label{3} 
  xG(x,Q^2)\,\,\propto\,\,\exp\{ \sqrt{\frac{4 N_c 
\as}{\pi}\,\ln\frac{Q^2}{Q^2_0}\,\ln\frac{1}{x}}\,\,\}\,\,.
\eeq

{\bf 2.}  In the region of $\ln \frac{1}{x}\,\approx\,\,\ln^2 Q^2$ we have
to 
take into account the correction to the DGLAP equation,mainly, related to 
so called leading log(1/x) approximation of perturbative QCD (LL(1/x)A). 
The LL(1/x)A have been studied by L.N.Lipatov and his collaborators 
\cite{BFKL}. We found that we can use for the asymptotic in our 
kinematic region the BFKL equation, but with one important additional 
ingredient: the running QCD coupling constant. We solve this equation 
within the accuracy that we needed ( in so called semiclassical 
approximation ). The result is that the gluon density is still rising in 
the region of low $x$ even more rapidly than in the DGLAP evolution.

{\bf 3.}  The increase of the gluon density leads to a new problem in
deeply 
inelastic scattering, namely the violation of $s$-channel unitarity, the 
requirement that the total cross section for virtual-photon absorption
be smaller than the size of a hadron
\beq  \label{4}
\s(\g^* N)\,\,\ll\,\,\pi R^2_h\,\,.
\eeq
Using even the DGLAP result ( see \eq{3} ) one can see that the unitarity 
constraint will be violated at at $x \,<\,x_{cr}$,
 where $\ln \frac{1}{x_{cr} }= c \ln^2Q^2$ and  $c$ 
is well defined constant. 
The resolution of this problem cannot be in the confinement phenomena. 
We must look for the origin and solution of this problem within 
perturbative QCD.

\begin{figure}
\centerline{\psfig{file= 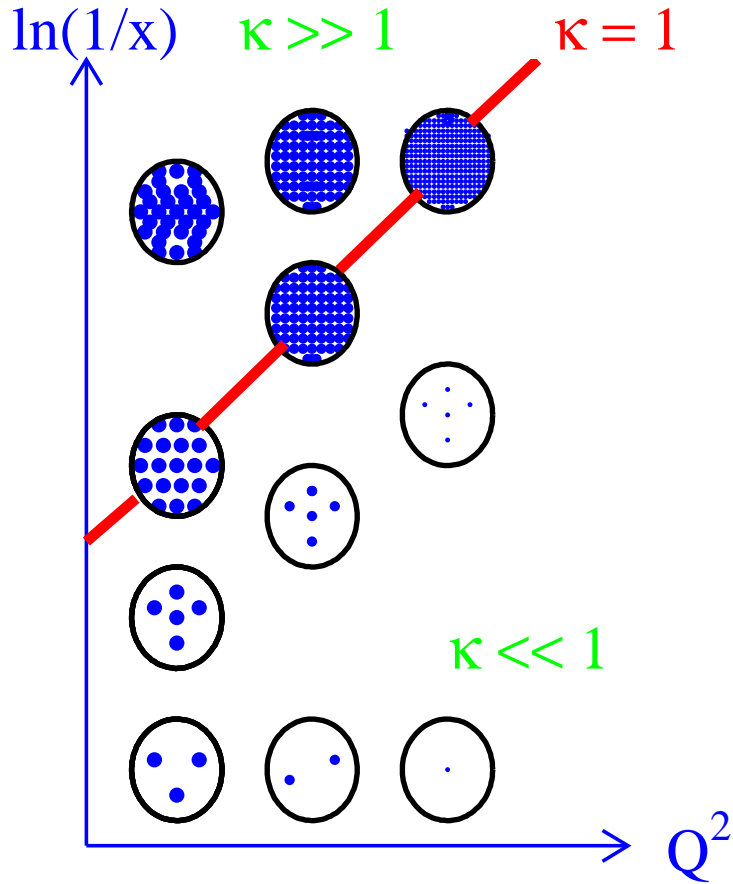,width=100mm}}
\caption{ \em Parton evolution in the transverse plane.}
\label{fig1}
\end{figure}

{\bf 4.}  We understood what happens in the region of small $x$ by
examining the 
parton distribution in the transverse plane (see Fig.1 ). Our probe feels
those 
partons with size $ r_p\,\sim \,\frac{1}{Q}$. At $x\,\approx\,1 $ a few 
parton are distributed in the hadron disc. If we choose $Q$ such that 
$r^2_p\,\ll\,R^2_h$ then the distance between partons in the transverse 
plane is much larger than their size, and we can neglect the interaction 
between partons. The only essential process is the emission of partons,  
which has been taken into account in QCD evolution. As $x$ decreases
for fixed $Q^2$ , the number of partons increases. and at value of 
$x=x_{cr}$, partons start to populate the whole hadron disc densely. 
For $x\,<\,x_{cr}$ the partons overlap spatially and begin to interact 
throughout the disc. For such small $x$ values, the processes of 
recombination and annihilation of partons should be as essential as 
their emission. However, neither process is incorporated into any 
evolution equation. What happens in the kinematic region 
$x\,<\,x_{cr}$ is anybody's guess. We suggested that parton density 
saturates, i.e. the parton density is constant in this domain.

{\bf 5.}  To take interaction and recombination of partons into account we
must 
identify a new small parameter that lets us estimate the accuracy of our 
calculation. We found this parameter:
\beq \label{5}
\kappa\,\,=\,\,x G( x,Q^2) \,\cdot\,\frac{\sigma(GG)}{\pi 
R^2_h}\,\,=\,\,\frac{N_c \as \pi}{2 Q^2 R^2_h} \,x G(x,Q^2)\,\,,
\eeq
where $\s$ is the cross section of gluon - gluon interaction and $R_h$ 
is the size of a hadron. The numerical factor in \eq{5} was evaluated by 
Mueller and Qiu \cite{MUQI}. This parameter $\kappa$ is the probability 
of a gluon recombination during the  cascade. The unitarity constraint 
can be rewritten in the form
\beq \label{6}
\kappa\,\,\leq\,\,1\,\,.
\eeq
We rewrote the amplitude of DIS as a perturbation series in this 
parameter which we resummed. The equation that we obtained can be 
easily understood by considering the structure of the QCD cascade in a 
fast hadron. Two processes occur inside the cascade:

\centerline{\it  Emission ($ 1 \,\rightarrow \,2$) with probability
\,$\propto 
\,\as \rho\,$;}

\centerline{\it Annihilation ($ 2 \,\rightarrow\,1$) with 
probability\,$\propto\, \as^2\,r^2_p \,\rho^2\,\propto\,\frac{\as^2}{Q^2}
\rho^2\,$;}
  
where $\rho$ is the density of gluons ( $\rho \,=\,\frac{x G(x,Q^2)}{\pi 
R^2_h})$. The number of partons in a phase space cell ($\Delta \Phi\,=\,
\Delta \ln(1/x)\,\Delta \ln Q^2$) increases due to emission and 
decreases due to  annihilation. Thus the balance equation reads:
\beq \label{7}
\frac{\partial^2 \,\rho}{\partial \ln\frac{1}{x} \,\partial\,\ln 
Q^2}\,\,=\,\,\frac{N_c \as}{\pi} \,\rho\,\,-\,\,\frac{\as^2 
\gamma}{Q^2}\,\rho^2\,\,,
\eeq
or in terms of the gluon structure function:
\beq \label{8}
\frac{\partial^2 \,x G(x,Q^2)}{\partial \ln\frac{1}{x} \,\partial\,\ln 
Q^2}\,\,=\,\,\frac{N_c \as}{\pi} \,x G(x, Q^2)\,\,-\,\,\frac{\as^2 
\gamma}{\pi\,Q^2\,R^2_h}\, ( x G(x,Q^2))^2\,\,.
\eeq
This is so-called the GLR equation. The factor $\gamma$ has been 
calculated by Mueller and Qiu and it is equal \cite{MUQI} 
$\gamma=\frac{N_c \pi^2}{8}$.

{\bf 6.}  We found the semiclassical solution to the GLR equation. It
turns out 
that this equation has a critical line:
\beq \label{9}
\ln \frac{1}{x_{cr}}\,\,=\,\,\frac{b}{32 N_c}\,\ln^2(Q^2/\Lambda^2)\,\,,
\eeq
where $\as(Q^2)\,=\,\frac{4 \pi}{b \ln (Q^2/\Lambda^2)}$.  For $x > 
x_{cr}$ , it suffices to find the solution of the linear DGLAP equation 
with the new boundary condition on the critical line. For $x < x_{cr}$ 
we
have a separate system of trajectories and solution does not depend on 
the solution for $x >x_{cr}$. We found  that the GLR equation is not 
the right tool to solve the problem in this region.

{ \bf 7} We realized that the critical line gives a new scale for the
value of 
the typical transverse momentum in the parton cascade, namely
\beq \label{10}
q^2_t\,\,=\,\,q^2_0(x)\,|_{x\,\rightarrow\,0}\,\rightarrow\,\,\Lambda^2 
\,e^{\sqrt{\frac{32 N_c}{b} \,\ln\frac{1}{x}}}\,\,.
\eeq
 This new scale plays a role of the infrared cutoff in all inclusive 
processes and leads to a number of prediction. The most important from 
them is the fact that the main contribution to the production processes 
gives the minijet production. We have studied these predictions in our paper 
during the past decade\cite{GLR2}.
\section{DIS at low $\mathbf x$\,\,( during the past 5 years ).}

During the last 5  years I have tried to understand better the relation 
between our approach and Wilson Operator Product Expansion. It turns out
( see Ref.\cite{LRS} ) that all high twist operators become important 
just
in the kinematic region near to the critical line. We can reformulate 
the problem of the low $x$ asymptotic as the problem to find the 
anomalous dimension for high twist operators. In Ref.  \cite{LRS} we 
found the anomalous dimension for twist four operator. We realized also 
that we made a mistake in our estimates of the contribution of the 
diagrams that killed the Pomeron approach (see also the paper of 
J.Bartels who did this first \cite{BAR}). The next step of my approach 
to the problem was to find the anomalous dimension for all high twist 
operators and to obtain a new evolution equation that replace the GLR 
one.  Eric Laenen and me found the solution to these two problems in our 
papers \cite{LL1}\cite{LL2}  as well as the solution to the new 
evolution equation. However, this new progress did not change the 
qualitative picture that I have described in the previous section.

We also developed the approach  how to take into account the new scale for 
the typical transverse momentum in our usual approach to hard processes 
based on factorization theorem \cite{FACTOR}. We  (M.G.Ryskin, 
A.G.Shuvaev,  Yu.M. Shabelsky and me ) proposed  so called the 
transverse momentum factorization \cite{LRSS} simultaneously with 
S.Catani,M, Ciafoloni and F.Hauptmann and J.Collins and R.K. Ellis 
\cite{TF}.  In this approach we introduce the unintegrated gluon 
structure function and the cross section of the interaction between 
partons off mass shell. Using these new ingredients we were able to 
write the convolution formulae in the analogous way with the usual 
approach. We proved that both of these new values can be calculated 
using the evolution equation at least in the leading log approximation 
of perturbative QCD.

Part of my activity  was devoted to better understanding of the evolution 
equations in the region of small $x$. The main questions that we 
approached were the interrelations between the DGLAP and BFKL evolutions 
\cite{MW} \cite{EKL}, the estimates for higher order corrections 
\cite{EKL}, the evolution equations for the diffractive dissociation 
processes \cite{LW} and the nonperturbative contribution for the BFKL 
equation stems from the infrared and ultraviolet renormalons 
\cite{LEREN}.

\section{DIS at low $\mathbf x$ (now ).}
\centerline{\it 1. My contribution:}

The intensive experimental study of the low $x$ behaviour of the deep
inelastic structure functions at HERA rises a number of questions to the 
theorists. Two of them, namely, where is the BFKL Pomeron and where is 
the SC, are under my close investigation. We are only in the beginning 
and what has been done is the new evolution equation which is able to 
describe the region of $x \,<\,x_{cr}$ \cite{LAST}. For fixed $\as$ this
equation can be written in terms of $\kappa$ ( see \eq{5} and has a form:
\beq \label{11}
\frac{\partial^2 \kappa(x,Q^2)}{\partial \ln\frac{1}{x} \partial \ln 
Q^2}\,\,+\,\,\frac{\partial \kappa}{\partial \ln 
\frac{1}{x}}\,\,=\,\,\frac{N_c\as}{\pi}\,\{ C\,\,+\,\,\ln 
\kappa\,\,+\,\,E_1 (\kappa)\,\}\,\,,
\eeq
where C is the Euler constant and $E_1$ is the exponential integral.
We solve this equation in semiclassical approximation and showed that 
the result gives weaker shadowing corrections than the GLR equation.
I am going to work on this problem in the nearest future and my main 
idea to resolve the difficulties that I have pointed out is to prove 
that the BFKL Pomeron is hidden under sufficiently strong SC while the SC 
theirselves were taken into account as the initial condition for the 
evolution equations. Much work is needed to clear up the situation.

\centerline{\it 2. Two different theoretical approaches:}
I would like to mention here that actually we have two theoretical
approach to the high parton density QCD.
To understand these two approaches we have to look at the picture of a
high energy interaction in the parton model (see Fig.2 ). In the parton
approach the
fast hadron decays into point-like particles ( partons )  long before 
( typical time $ \tau \,\propto\,\frac{E}{\mu^2}$ ) the interaction with
the
target. However, during this time $\tau $,   all partons  are in the
coherent state which can be described by means of a wave function. The
interaction of the slowest ($``wee"$) parton with the target completely
destroys
 the coherence of the partonic wave function.
\begin{figure}
\centerline{\psfig{file=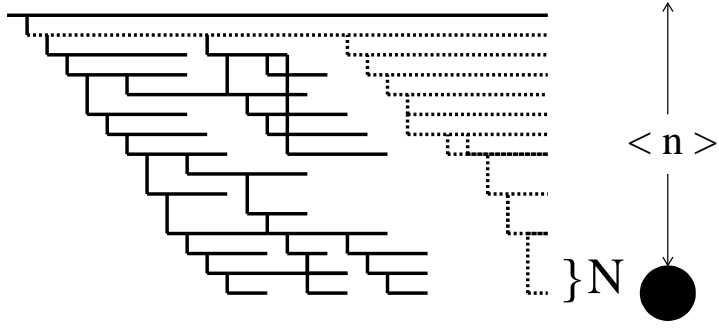,width=100mm}}
\caption{ \em The high energy interaction in the parton model.}
\label{fig2}
\end{figure}
The total cross section of such
an interaction is equal to
\begin{equation}
\label{12}
\sigma_{tot}\,\,=\,\,N\,\times\,\sigma_0
\end{equation}
where
\begin{itemize}
\item  $N$\,\,=
{ \it flux ( }$\mathbf renormalized$ {\bf ?!} { \it ) of ``wee"
partons\,\,;}
\item
$\sigma_0$\,\,=
{ \it the cross section of the
interaction of one
``wee" parton with the target\,\,.}
\end{itemize}
One can see directly from Fig.1 that the number of ``wee" partons is
rather large and it is equal to
\beq \label{13}
N\,\,\propto\,\,e^{ < n>}\,
=\,\frac{1}{x^{\omega_0}}\,\,\,\,with\,\,\,\,
 < n >\,=\,\omega_0
\,\ln(1/x)\,\,;\,\,\omega_0\,=C \alpha_S\,\,.
\eeq 
We have to renormalize the flux of ``wee" partons, since the total
cross section is the number of interactions  and if one has  several
``wee" partons with the same momenta they only give  rise to  one
interaction. 

If $ N\,\,\approx\,\,1$, we expect that the renormalization of the flux
will be small, and we use an approach with the following typical
ingredients:

$ \bullet$  Parton Approach;
$ \bullet$ Shadowing Corrections;
$ \bullet$ Glauber Approach;
$ \bullet$ Reggeon-like Technique;
$ \bullet$ AGK cutting rules\,.

However, when
$ N\,\,\gg\,\,1$,  we have to change our approach completely from  the
parton cascade to one based on semiclassical field approach, since due to
the uncertainty principle
 $\Delta N \Delta \phi \,\approx\,1$, we can consider the phase as a small
parameter.
Therefore, in this kinematic region our magic words are:

$ \bullet$   Semi-classical gluon fields;
$ \bullet$  Wiezs$\ddot{a}$cker-Williams approximation;
$ \bullet$ Effective Lagrangian for hdQCD;
$ \bullet$  Renormalization Wilson group Approach.

It is clear, that for $N \,\approx\,\,1$ the most natural way is to
approach the hdQCD looking for corrections to the perturbative parton
cascade. In this approach the pQCD evolution has been naturally included,
and it aims to describe the transition region. The key problem
 is to penetrate into the hdQCD region where $\kappa $ is large.
Let us call this approach ``pQCD motivated approach ". Namely, in this
approach we obtained \eq{11} which we consider as a correct tool to
evaluate a high parton density effects in DIS.

For $N \,\gg\,\,1$, the most natural way of doing is to use the effective
Lagrangian approach,  and remarkable progress has been achieved both in
writing of the explicit form  of this effective Lagrangian,  and in
understanding physics behind it  \cite{EL}. The key problem
for this approach was to find a correspondence  with pQCD. This problem
has been
solved \cite{KOV}.

Fig.3 shows the current situation on the frontier line in the offensive on
hdQCD.

\begin{figure}
\centerline{\psfig{file= 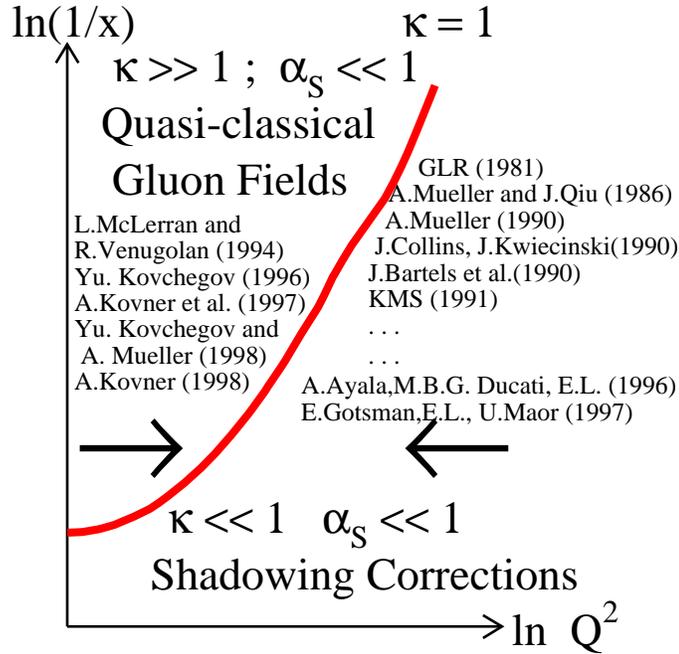,width=100mm}}
\caption{ \em The current situation in the battle for hdQCD.}
\label{fig3}
\end{figure}

Much more work has to be done before we will be able to find a solid
theoretical description of the kinematic region of hdQCD. The ``hot"
problem is to find a matching between two different approaches. The first
step has been done in this direction: in Ref. \cite{KOVL} was proven that
the effective Lagrangian approach gives the GLR equation for the limit of
sufficiently small parton densities. However, the interrelation between
\eq{11} and the equation suggested in Ref. \cite{KOVL} is still unclear.

\section{Conclusion.}
Not only me but many people are working on the problem of the Pomeron 
structure. My main idea was and is to look on this problem { \it going
from 
inside of the hadron}  or in other words { \it from small distances}.
Going in this direction we always feel the support of  perturbative 
QCD which allows us to check our imagination and to foresee the 
direction of  each new 
step in our attempts to solve the Pomeron problem. What is the Pomeron 
is still an open question and it is a challenge for all physicists who 
like to solve a difficult problem. The school of high energy 
phenomenology and the perturbative calculation created by Prof. Gribov 
in the theory  department of St.Petersburg (Leninhgrad) Nuclear Physics 
Institute and pushed forward
by my generation of physicists 
made this  department well known through all the world. The problem of the
Pomeron structure, even has not been solved yet,  gives a good training
for 
learning of most of the  secrets of the perturbative QCD approach and for
the 
improving of the calculation skill. It is a window to all difficult 
problems of QCD and any field theory approach. I am firmly believe that 
the new ideas are needed to make a progress. I ask a young theorist 
 to look back on my life with the Pomeron and admit 
that it was not so bad, at least, it was and is rather interesting life
without any boring situation, with many ups and downs, but always with 
some perspective for further progress.

\section{My several words about Prof. Gribov.}
Of course, I was a lucky guy having such a great teacher as Prof. Gribov,
who was one of the most outstanding physicists in this century. I was
lucky twice, since I had a three year experience of doing physics with
him. Formally speaking, we were doing several papers on the Reggeon -
Pomeron interactions. Really, we tried to get a picture of high energy
interaction based on the most advanced theory of that time. It was a hard
work and I learned several lessons which I follow during my life in
physics  and which I want to
share with you.
\begin{enumerate}
\item { \it Physics - first.} He always tried to understand the result
without mathematics, he did not trust the formal derivation of anything,
he wanted to understand what physics were used in the calculation. The
best for him was to obtain a result without calculation. Each day of
working with him we started with the question: ``What is the small
parameter in our approach?".   He called this process to create a picture.
He was really unhappy when he could not create such a picture, in which
everything should be selfconsistent and should not contradict any
theoretical result or/and any experimental data.

 Let me illustrate my
point telling you one story with me. It was a long ago at that time when 
we thought that there were only Reggeons and nothing except Reggeons.
I found a paper with the data of Lindenbaum where was a dip in
$t$-dependence in some reaction. I was young and remembered
everything that was thought in
the university. I came to Prof. Gribov, showed him the data and told him
that from my point of view it was very natural to have such a dip because
of diffraction.  After that I asked a naive question:  ``Why we cannot
describe such a simple physical phenomena in the Reggeon approach." 
He answered me something. I forgot what. However, he understood that he
failed to convince me. After that, during the next  two years, he 
was always 
coming
back to my question and, finally, when he understood that
 shadowing corrections or exchange
of
 Pomerons   were unavoidable ingredients of our theory he was
happy to explain me, how we should describe diffraction in the Pomeron
approach. 

 He could not live without selfconsistent picture. He spent the
past two decades in a some sort of isolation only because he could not
find a physical picture for such an important physical phenomena as
confinement of quarks and gluons. Unfortunately, he went away before
explaining us what picture he has finally found. Unfortunately, it is
almost impossible to understand everything from his paper. Could ancient
Greek, the smartest one, understand one of our paper translated to the
 perfect Greek
language of his time? I think, he could not. I feel as this Greek talking
with Prof. Gribov. You have to live with physical problem, to nurse it and
to think about the problem as he did to understand his picture. 

\item {\it Mathematics is only a tool.} The knowledge of mathematics was
requited but there was no a slight respect to it. Mathematics is only a
tool like a hammer to nail something. 

Once more, a story with me. It was an equation which even we did not put
on the paper. He gave a solution of it asking me  to check if everything
is
correct. I took a couple of sheets of paper covered by formulae. To my
surprise, I immediately found that the first line was wrong. My lord, the
second also turns out to be incorrect, the third, the fourth.... So
everything was wrong but the answer was absolutely correct! Next day, I
was
very proud and came with a pile of papers with correct solution.
The most instructive for me was his reaction. He said:``Genya, why you
spent so much time recovering how to get this solution. Why you did not
check, that this is a solution, directly substituting my formula into 
equation?".  For him it did not matter how you got the solution of
physical problem. The only thing, that counted, was to get a solution. 

\item {\it Physics is the experimental science.} Prof.Gribov was proud
being theorist because a theorist has  a privilege to know all
experiments and to create out of them a physical picture.  
He was always saying that theory costs nothing if it contradicts a single
experimental fact.

 He was upset when the first data from Srpukhov
accelerator came showing that there was a large difference between $K^{+}
p$ and $K^{-} p$ total cross sections. He sent me to ITEP to discuss with
Prof. Okun' and Prof. Ter-Martirosian these new data which seemed to
contradict the Pomeranchuk theorem. He was unhappy until the example was
built that the Pomeranchuk theorem could be violated without any harm for
theory. 

\item {\it Two duties.} Prof. Gribov was a man who enjoyed a freedom of
thinking and we had to follow only two  rules that restricted our freedom.
Any member of our
department has
only two duties: 
\begin{enumerate}
\item To attend any seminar in our department or in other places  that you
visit\,; 

\item To answer any question that an experimentalist ask you.
\end{enumerate}

Our seminars were the place where we worked. They lasted as much time as
we needed to understand a problem. Sometimes we needed several days to do
this. Prof. Gribov was the leader, he understood better and attacked a
poor speaker first. A speaker had to  have the  guts to stay against these
attacks and, in spite of all obstacles, he had to try to defend himself.
The tactics was very simple: try to formulate during the first ten
minutes   the physical problem that you solved and the result. If Prof.
Gribov would understand your problem,  he would help you to
find a solution to  this
problem. Of course, he would not listen to you for the first one hour
since he would try to solve a problem by himself. Sometimes, he did this
and after that  you were a lucky guy ,if his solution coincided with
yours.
If not, he would try to convince you that his solution was better.
Sometimes, he would not been able to find a solution. In this case, you
were in a good situation. You could tell Prof. Gribov something new and
got
his full respect.

\item {\it The full responsibility.}
You are responsible for any quoted result as it is your own. Presenting
your talk in the seminar,  you have to answer any question on the subject
of
you talk, especially, if you referred to some results of others.
You cannot tell that this theorem was proven by somebody. You have to 
present the proof which will be scrutinized  together with your ability
to understand your subject. The rules were simple - no publication without
presenting your paper in our seminar.

Unfortunately, the culture of such working seminars is passing away. 
Outside Russia I met only in two places the same atmosphere as was in
Gribov's seminars: in our Tel Aviv  and in Minnesota Universities, but
even there the seminar is restricted by one hour. 
\end{enumerate}

I grew up as a physicist under strong influence of Prof. Gribov. I am
afraid that for the rest  of my life I will look  at physics with his
eyes. We lost him too earlier to cope with this fact. I hope, that he will 
live in us. This is an idea behind this report on my own way to approach
his beloved problem - the Pomeron. 

{\bf Acknowledgements:}
I am very grateful to my friends A. Kaidalov, O. Kancheli and M. Ryskin
for a life long discussion of the Pomeron. I thank all my collaborators
who worked  and suffered with me on this difficult problem  and who have  
not
lost their optimism. My special thanks go to E. Gotsman and U. Maor who
started to re-analyze with me what we know about
the Pomeron and their interactions  and supported me in all my crazy ideas
on the subject. 

This paper was completed at DESY and I would like to thank all members of
the Theory group at DESY for their hospitality.

I also express my deep gratitude to everybody who made possible this 
`Gribov Memorial Volume'. Hope , that memory of Prof. Gribov will
stimulate
us to do better physics, the best that we can.


\begin{thebibliography}{99}
\bibitem{FRST}
M. Froissart: {\it Phys. Rev} {\bf 123} (1961) 1053.
\bibitem{MAR}
A. Martin: {\it ``Scattering Theory: Unitarity, Analyticity and 
Crossing."},Lectures Notes in Physics, Springer - Verlag, 1969.
\bibitem{GRIB1}
V.N. Gribov: {\it Sov. Phys. JETP} {\bf 53} (1967) 654.
\bibitem{MG}
V.N. Gribov and A.A. Migdal:  {\it Sov. Phys. JETP} {\bf 55} (1968) 
1498; {\it Sov.J.Nucl.Phys.} {\bf 8} (1968) 1002.
\bibitem{GML1}
V.N.Gribov,E.M.Levin and A.A.Migdal: {\it Sov.J.Nucl.Phys.} {\bf 11} 
(1970) 673;{\bf 12} (1970) 173;{\it Sov.Phys. JETP} {\bf 32} (1970)
1158.
\bibitem{LKS}
V.A.Kudryavtzev, E.M.Levin and A.A.Schipakin: {\it Sov.J.Nucl.Phys.} {\bf 
9} 
(1969) 1274;{\bf 10} (1969) 1748;{\bf 11} (1970)858;\\
V.A.Kudryavtzev and E.M.Levin: {\it Sov.J.Nucl.Phys.} {\bf 
18} (1973) 451.
\bibitem{K}
Ya.I.Azimov, E.M. Levin,M.G.Ryskin and V.A. Khoze:{\it Nucl.Phys.} {\bf 
B89} (1975) 508;{\it Sov.J.Nucl.Phys.} {\bf 
21} 
(1975) 413;{\bf 23} (1976) 853;\\
Ya.I.Azimov, E.M. Levin,M.G.Ryskin,M.I.Strikman
 and V.A. Khoze:{\it Sov. Phys. JETP Lett.} {\bf 
23} 
(1976) 121;\\
E.M.Levin,M.G.Ryskin,M.I.Strikman and G.G.Takhtamyshev:
{\it Nucl. Phys.} {\bf B123} (1977) 1020.
\bibitem{KREV}
Ya.I.Azimov, E.M. Levin,M.G.Ryskin and V.A. Khoze:IX Winter Leningrad 
School, v.II,p.5, 1974.
\bibitem{GLM}
E. Gotsman, E.M. Levin and U. Maor: {\it Z. Phys} {\bf C57} (1993) 667;
{\it Phys. Lett.} {\bf B309} (1993) 109; {\bf B347} (1995) 424; {\it 
Phys. Rev.} {\bf D49} (1994) R4321.
\bibitem{LN}
F. Low:{\it Phys. Rev.} {\bf D12} (1975) 163; \,\,S. Nussinov: {\it Phys.
Rev. Lett.} {\bf 34} (1975) 1286, {\it Phys. REv.} {\bf D14} (1976) 244. 
\bibitem{AFS}
D. Amati, S. Fubini and A. Stangellini:{\it Nuovo Cim.} {\bf 26} (1962) 
826.
\bibitem{TER}
M. Baker and K.A. Ter-Martirosyan: {\it Phys. Rep.} {\bf 28C} (1976) 3 
and
references therein.
\bibitem{VEN}
G. Veneziano:{\it Nouvo Cim.} {\bf 57A} (1968) 190; {\it Phys. Rep.} 
{\bf 9} ( 1974) 199.
\bibitem{MM1}
E.M. Levin and M.G. Ryskin: {\it Phys. Lett.} {\bf B41} (1972) 681;
 {\it Sov. Phys. JETP Lett.} {\bf 16} (1972) 495; {\bf 17} (1973) 669;
{\it Sov. J. Nucl. Phys.} {\bf 17} (1973) 388; {\bf 18} (1973) 431,1108.
\bibitem{AGK}
V.A. Abramovsky, V.N. Gribov and O.V. Kancheli: {\it Sov. J. Nucl. Phys.}
{\bf 18} (1973) 308.
\bibitem{MM2}
E.M. Levin and M.G. Ryskin: 
 {\it Sov. Phys. JETP Lett.} {\bf 17} (1973) 669; {\bf 18} (1973) 654;
{\it Sov. J. Nucl. Phys.} {\bf 19} (1974) 389,669 904; {\bf 20} (1974) 
519; {\bf 21} (1975) 352,1072,1281; {\bf 22} (1975) 428;{\bf 23} (1976) 
423; {\bf 24} (1976)640.
\bibitem{CK}
A.Capella, A. Krzywicki and  E.M. Levin: {\it Phys. Rev.} {\bf D44} 
(1991) 704.
\bibitem{FE}
R.P. Feyman: {\it Phys. Rev. Lett.} {\bf 23} (1969) 1415; 
{``Photon-Hadron Interaction"} N.Y. Benjamin, 1972.
\bibitem{BJ}
J.D. Bjorken: {\it``Proceedings of the Int. Symposium on Electron and 
Photon Interaction at High Energy"} p.281,Cornell,1971 and references 
therein.
\bibitem{GR}
V.N. Gribov: {\it Sov.J.Nucl.Phys} {\bf 9} (1969) 640; {``Proc. VII LNPI 
Winter School"} v.II,p.5, Leningrad 1973.
\bibitem{P1}
E.M. Levin and M.G. Ryskin: {\it Sov. Phys. JETP} {\bf 69} (1975) 412.
\bibitem{P2}
E.M. Levin and M.G. Ryskin: {\it Sov.J.Nucl.Phys.} {\bf 19} (1974) 519, 
{\bf 22} (1975) 428.
\bibitem{P3}
E.M. Levin and M.G. Ryskin: {\it Sov.J.Nucl.Phys.} {\bf 25} (1977) 349.
\bibitem{P4}
E.M. Levin and M.G. Ryskin: {\it Sov.J.Nucl.Phys.} {\bf 27} (1978) 
790,{\bf 29} 1311.
\bibitem{P6}
E.M. Levin and M.G. Ryskin: {\it Sov.J.Nucl.Phys.} {\bf 31} (1980) 429.
\bibitem{TTWU}
B.M. McCoy and T.T. Wu: {\it Phys. Rev.} {\bf D 12} (1975) 546,577.
\bibitem{MS}
S.G. Matinyan and A.G. Sedrokyan: {\it Sov.Phys.JETP lett.} {\bf 23} 
(1976) 588, {\bf 24} (1976) 240, {\it Sov.J.Nucl.Phys.} {\bf 24} (1976) 
844.
\bibitem{GLR1}
L.V. Gribov, E.M. Levin and M.G. Ryskin:{\it Sov.Phys,JETP} {\bf 80} 
(1981) 185, {\it Phys. Lett.} {\bf B100} (1981) 173,{\bf B101} (1981) 
185,{B121} (1983)65,
{\it Nucl.Phys.} {\bf B188} (1981) 555, {\it Sov.J.Nucl.Phys.} {\bf 
35} (1981) 1278.
\bibitem{GLRREV1}
E.M. Levin and M.G. Ryskin: {\it Phys. Rep.} {\bf 100} (1983) 1.
\bibitem{LRREV}
E.M. Levin and M.G. Ryskin: {\it Phys. Rep.} {\bf 189} (1990) 267.
\bibitem{LLREV}
E. Laenen and E. Levin: {\it Ann. Rev. Nucl. Part.Sci} {\bf 44} (1994) 
199.
\bibitem{GLAP}
V.N. Gribov and L.N. Lipatov:  {\it Sov. J. Nucl. Phys.} {\bf 
15} (1972) 438;\,\,
 L.N. Lipatov: {\it Yad. Fiz.} {\bf 20} (1974) 181;\,\,
G. Altarelli and G. Parisi:{\it Nucl. Phys.} {\bf B 126} (1977) 298;\,\,
Yu.L.Dokshitzer: {\it Sov.Phys. JETP} {\bf 46} (1977) 641.
\bibitem{BFKL}
 E.A. Kuraev,  L.N. Lipatov and V.S. Fadin: {\it Sov. Phys. JETP} {\bf 45}
        (1977) 199 ;
\,\,Ya.Ya. Balitskii and L.V. Lipatov:{\it Sov. J. Nucl. Phys.} 
{\bf 28} (1978)
822;\,\,L.N. Lipatov: {\it Sov. Phys. JETP} {\bf 63} (1986) 904.
\bibitem{MUQI}
A.H. Mueller and J. Qiu: {\it Nucl. Phys.}  {\bf B268} (1986) 427.
\bibitem{GLR2}
E.M. Levin and M.G. Ryskin: {\it Sov.J.Nucl.Phys.} {\bf 45} (1987) 
234,{\bf 47} (1988) 1398,{\bf 50} (1989) 881,{\bf 53} (1991) 653, {\it 
Nucl.Phys.} {\bf B304} (1989) 805.
\bibitem{LRS}
E.M. Levin, M.G. Ryskin and A.G. Shuvaev: {\it Nucl.Phys.} {\bf B387} 
(1992) 589.
\bibitem{BAR}
J.Bartels: {\it Phys.Lett.} {\bf B298} (1993) 204, {\it Z.Phys.} {\bf 
C60} (1993) 471.
\bibitem{LL1}
E.Laenen, E. Levin and A.G. Shuvaev: {\it Nucl. Phys.} {\bf B419} (1994) 
139.
\bibitem{LL2}
E. Laenen and E. Levin: {\it Nucl. Phys.} {\bf B451} (1995) 207.
\bibitem{FACTOR}
J.C.Collins, D.E. Soper and G. Sterman: {\it Nucl.Phys.} {\bf B308} 
(1988) 833.
\bibitem{LRSS}
E.M. Levin, M.G. Ryskin, Yu.M. Shabelsky and A.G. Suvaev: {\it 
Sov.J.Nucl.Phys.} {\bf 53} (1991) 653.
\bibitem{TF}
S.Catani,M.Ciafaloni and F.Hauptmann: {\it Phys.Lett.}{\bf B242} (1990) 
97,{\it Nucl.Phys.} {\bf B366} (1991) 135;\\
J. C. Collins and R.K. Ellis: {\it Nucl.Phys.} {\bf B360} (1991) 3.
\bibitem{MW}
G. Marchesini, E.M. Levin, M.G.Ryskin and B.R. Webber: {\it Nucl.Phys.}
{\bf B357} (1991) 167.
\bibitem{EKL}
R.K. Ellis, E.M. Levin and Z. Kunszt: {\it Nucl.Phys.} {\bf B420} (1994) 
517.
\bibitem{LW}
E.M. Levin and M. Wuesthoff: {\it Phys. Rev.} {\bf D50} (1994) 4306.
\bibitem{LEREN}
E. Levin: {\it Nucl.Phys.} {\bf B453} (1995) 303.
\bibitem{LAST}
A.L. Ayala, M.B. Gay Ducati and E.M. Levin: 
{\it Nucl.  Phys. } {\bf B511} ( 1998 ) 355, {\bf B493} ( 1997 ) 305. 
\bibitem{EL}
L. McLerran and R. Venugopalan: {\it Phys. Rev.} {\bf D49} (1994)
2233,3352, {\bf D50} (1994) 2225, {\bf D53} (1996) 458;
 J. Jalilian-Marian, A. Kovner, A. Leonidov and H. Weigert:{ hep -
ph/9701284}{ \tt hep-ph/9706377};  J.
Jalilian-Marian, A. Kovner, L. McLerran  and H. Weigert: { \it Phys.
Rev.}{ \bf D55} (1997) 5414; Yu. Kovchegov: {\it Phys. Rev.}{ \bf D54}   
(1996) 5463; Yu. Kovchegov and A.H. Mueller: {\tt hep-ph/9802440}.
\bibitem{KOV}
 J. Jalilian-Marian, A. Kovner, A. Leonidov and H. Weigert:  {\tt hep -
ph/9701284},\,\,{ \tt hep-ph/9706377}.
\bibitem{KOVL}
J. Jalilian-Marian, A. Kovner, A. Leonidov and H. Weigert:{ \tt
hep-ph/9807462}. 

\end{thebibliography}
\end{document}